\documentclass[twocolumn]{extarticle}  
\usepackage[utf8]{inputenc}
\usepackage[T1]{fontenc}

\usepackage{geometry}
\usepackage{fancyhdr} 
\usepackage{graphicx}

\usepackage{hyperref}
\hypersetup{colorlinks=true,linkcolor=blue,urlcolor=blue,citecolor=blue}

\usepackage{amsmath,amssymb,mathrsfs,mathtools}

\newcommand{\lb}{[\![}
\newcommand{\rb}{]\!]}

\usepackage{centernot}

\numberwithin{equation}{section} 

\usepackage{kpfonts}

\usepackage[affil-it]{authblk}

\usepackage{abstract}

\usepackage{titlesec}
\titleformat*{\section}{\large\bfseries}

\usepackage[labelfont={bf},textfont={small},width=230pt]{caption}

\usepackage{cite}
\title{ \huge Deconjugation of canonical variables and the Koopman-von Neumann theory 
}

\author[1]{Mustafa Amin \thanks{\href{mailto:m.amin@uleth.ca}{m.amin@uleth.ca}}}
\author[2]{Mark A.~Walton \thanks{\href{mailto:walton@uleth.ca}{walton@uleth.ca}}}

\affil[1,2]{ 
    Department of Physics \& Astronomy, University of Lethbridge, Lethbridge, Alberta, T1K 3M4, Canada
}

\date{}

\begin{document}

\twocolumn[  
    \begin{@twocolumnfalse}  

\maketitle

\begin{abstract}
    The Koopman-von Neumann (KvN) theory is one where the dynamical momentum is not canonically conjugate to position, i.e., position and momentum are deconjugated.  From this point of view, we show that the KvN theory arises from quantum mechanics, extracting classical equations of motion from quantum ones.  However, preserving the canonical structure of the theory requires introducing ``auxiliary'' canonical conjugates to position and momentum.  We show that using the KvN formulation to study the interaction between quantum and classical systems forces the auxiliary variables to take on a physical role.  While giving rise to classical behaviour, the KvN theory might be more than classical.

\vspace{1cm}

\end{abstract}

    \end{@twocolumnfalse}  
]  

\saythanks  

\section*{Introduction}

This paper grew out of an attempt to understand the underlying structure of the Koopman-von Neumann (KvN) formulation of classical mechanics.  In that formulation, physical quantities are represented as operators acting on Hilbert space, much like the operator formulation of quantum mechanics.  However, a tension arises between the classical nature of the theory and the canonical structure of operator dynamics.  On one hand, the physical operators must commute in order to give classical vanishing uncertainty relations.  On the other hand, the natural dynamical bracket for operators is the commutator, which would identically vanish for the classical physical operators.  Consistency then requires the introduction of auxiliary variables that are canonically conjugate to the physical position and momentum, but are typically considered unphysical.

Unlike canonical theories, the KvN formulation then contains an object that is conjugate to the physical position, but is not itself the physical momentum.  Similarly, there exists an object that is conjugate to the physical momentum, but is not the physical position.  This inspires the separation of the concepts of \textit{conjugate momentum} and \textit{dynamical momentum}, which are usually identified as one and the same.  


Here we examine the process of \textit{deconjugating} the dynamical momentum from its associated position in quantum mechanics.  The analysis presented here shows that the KvN theory of a classical system can be ``extracted'' from its quantum analogue in a manner akin to, but not the same as, taking the classical limit.  In this respect, we find hints that the auxiliary variables of the theory might not be easily dismissed as unphysical.  The KvN formulation seems to describe more than just classical mechanics.

The possibility that the auxiliary variables are physical is reinforced when we let a KvN system interact with a quantum one.  If the KvN formulation is used to describe quantum-classical interactions, we show that the extra variables show up in the equations of motion of the manifestly physical ones.  The mixing of physical and auxiliary variables is clearly demonstrated in the Heisenberg picture.

This paper is organized as follows.  We set up some useful notation in Sec.~\ref{sec:notation} and a summary of the canonical formulation in Sec.~\ref{sec:summary-hamiltonian}.  We then introduce and discuss the process of deconjugation in Sec.~\ref{sec:deconj} along with the resulting equations of motion.  The KvN formulation and its relationship to the classical limit are discussed in Sec.~\ref{sec:kvn} through the lens of deconjugation.  Finally, we show in Sec.~\ref{sec:no-part-deconj} the role of the auxiliary variables in the physical equations of motion of a classical-quantum system.

\section{Notation}\label{sec:notation}

Square brackets $[u,v]$ are used to denote the commutator of two variables $uv-vu$.  Double square brackets denote the commutator divided by $i\hbar$
\begin{align}
    \lb u,v \rb = \frac{1}{i\hbar}[u,v]~.
\end{align}
This notation is useful for comparison with classical mechanics.  Note that, for variables $u$ and $v$ that are not explicit functions of $\hbar$, it can be shown that
\begin{align}
    [u,v] = O(\hbar) &\implies \lim_{\hbar\to0}[u,v] = 0~, \\
    \lb u,v \rb = O(\hbar^0) &\implies \lim_{\hbar\to0}\lb u,v \rb \neq 0~.
\end{align}

We will use an underline to signify an expression where the fundamental variables $q$ and $p$ commute
\begin{align}
    qp \neq pq~, \quad \underline{qp} = \underline{pq}~.
\end{align}
This implies
\begin{align}
    \underline{f(q,p)} = f(q,p)\big{|}_{qp=pq}~.
\end{align}
Note that a vanishing expression implies the vanishing of its underline, but not the other way around
\begin{align}
    f(q,p) = 0 &\implies \underline{f(q,p)} = 0~,\\
    \underline{f(q,p)} = 0 &\centernot\implies f(q,p) = 0~.
\end{align}

Finally, in Sec.~\ref{sec:deconj}, we will use the symbols $\tilde{q}$, $\tilde{p}$ and $\tilde{H}$ that correspond to the usual $-\lambda_p$, $\lambda_q$ and the Liouvillian $L_H$ in the literature on KvN theory.  These variables will be discussed in context.  Any variable $u$ should be understood as a function $u(q,p)$ of $q$ and $p$ only, while a tilde over a symbol $\tilde{u}$ implies that it is a function $\tilde{u}(q,\tilde{p},\tilde{q},p)$~.

\section{Summary of the Hamiltonian formulation}
\label{sec:summary-hamiltonian}

In a Hamiltonian theory, one usually starts with pairs of variables $(q,p)$, a Hamiltonian $H(q,p)$ and a dynamical bracket $\lb\cdot\,,\cdot\rb$ with which a set of fundamental relations are defined
\begin{align}\label{eq:fund-rel-1}
    \lb q,p\rb  = 1~,\quad \lb q,q\rb =\lb p,p\rb =0~.
\end{align}
The dynamical bracket $\lb \cdot\,,\cdot\rb $ is a binary operation between the dynamical variables $(u,v,w,\cdots)$ that obeys the following conditions:
\begin{align}
    \lb v,u\rb  &= -\lb u,v\rb~, \\
    \lb u+v,w\rb  &= \lb u,w\rb  + \lb v,w\rb~, \\
    \lb  \lb u,v\rb ,w \rb  &= \lb  \lb u,w\rb ,v \rb  + \lb  u,\lb v,w\rb  \rb~, \\
    \lb uv,w\rb  &= \lb u,w\rb v + u\lb v,w\rb ~.
\end{align}
Those are the antisymmetry, linearity, Jacobi and Leibniz properties.  In addition, the bracket vanishes between dynamical variables and scalars $\lb u,c\rb =0$~.  Using the fundamental relations and the Leibniz property, it is easy to show that the bracket acts like a derivative (e.g., $\lb q^n, p \rb = nq^{n-1}$), so we can define
\begin{align}
    \partial_q u &:= \lb u,p \rb~,\\
    \partial_p u &:= \lb q,u \rb~
\end{align}
even for noncommutative variables.

The Hamiltonian $H$ defines the time evolution of the system.  In the Heisenberg picture, the system evolves through the dynamical variables
\begin{align}
    \frac{d}{dt} u = \lb u,H\rb  + \frac{\partial}{\partial t} u~,
\end{align}
while in the Schrödinger picture, the dynamical variables are static and the system evolves through its state $\rho$ according to the von Neumann equation
\begin{align}\label{eq:rho-eq}
    \lb \rho,H\rb  + \frac{\partial}{\partial t} \rho = 0~.
\end{align}

The Heisenberg and Schrödinger pictures are related through the equality of the time dependence of expectations values in a given state $\rho$
\begin{align}
    \langle u(t) \rangle_\rho = \langle u \rangle_{\rho(t)}~.
\end{align}
The left hand side of the above equation is calculated in the Heisenberg picture and the right hand side is in the Schrödinger picture.  Using either pictures, the Ehrenfest equation can be derived as
\begin{align}\label{eq:Ehrenfest}
    \frac{d}{dt} \langle u \rangle = \left\langle \lb u,H \rb \right\rangle + \left\langle \frac{\partial}{\partial t} u \right\rangle~.
\end{align}
This equation for the evolution of expectation values is a consequence of the Hamiltonian formulation and the definition of expectation values.

It is clear that a central role is played by the algebraic properties of the dynamical bracket along with the fundamental relations.  In fact, none of the above \textit{formal structure} is specific to either classical or quantum dynamics.  The same formal equations hold for classical mechanics replacing the re-scaled commutator $\lb\cdot\,,\cdot\rb$ with the Poisson bracket $\{\cdot\,,\cdot\}$ and the state $\rho$ would be the Liouville phase-space distribution.  Of course, in a classical theory, all variables commute.  Further, the formal structure above is also valid for phase-space quantum mechanics with the appropriate identification of variables and their product.

\section{Deconjugation}
\label{sec:deconj}

Using the fundamental relations stated in the previous section, we can show that the value of the dynamical variable $u$ after a translation $q \to q+\epsilon$ is given by
\begin{align}\label{eq:taylor}
    u(q+\epsilon,p) = e^{\epsilon\lb \cdot\,,p \rb} u(q,p)~.
\end{align}
That is, because $p$ is the \textit{canonical conjugate} of $q$ (i.e., $\lb q,p \rb=1$), it is the generator of translation along $q$.  The role of the conjugate is independent of the dynamics (equations of motion) and the Hamiltonian.

On the other hand, the Heisenberg equation of motion for $q$ relates $\dot{q}$ and $p$
\begin{align}\label{eq:relate-qdynp}
    \dot{q} = \lb q,H \rb = v(q,p)~.
\end{align}
The dot indicates a total time derivative.  For example, the simple Hamiltonian $H = p^2/2m + V(q)$ gives $\dot{q}=p/m$~.  It is the appearance of $p$ in the Hamiltonian that lands it on the right hand side of~\eqref{eq:relate-qdynp}~.  The role played by the variable $p$ in this equation is that of \textit{dynamical momentum}.  We call it ``dynamical'' because it arises from the equations of motion which depend on the form of the Hamiltonian.

Conjugate momentum and dynamical momentum are different concepts and need not be identical.  To demonstrate the difference, consider a system of two degrees of freedom with the usual fundamental relations (with $i,j \in {1,2}$)
\begin{align}
    \lb q_i\,p_j \rb = \delta_{ij}~, \quad \lb q_i,q_j \rb = \lb p_i,p_j \rb = 0~.
\end{align}
Let the Hamiltonian be given by
\begin{align}
    H = \frac{p_1 p_2}{m} + k \, q_1 q_2~,
\end{align}
then the equations of motion are
\begin{align}
    \dot{q}_1 &= p_2/m~, \quad \dot{q}_2 = p_1/m~,\\
    \dot{p}_1 &= -k \, q_2~, \quad \dot{p}_2 = -k \, q_1~.
\end{align}
In this example, we see that $p_1$ is the conjugate momentum of $q_1$ while its dynamical momentum is $p_2$~.  The Heisenberg equations of motion were used here because they provide a straightforward illustration of the different concepts of conjugate and dynamical momentum.  That distinction, however, is not specific to the Heisenberg picture.  Observe that the dynamical momentum can be defined as the one resulting from the bracket $\lb q, H \rb = v(q,p)$ without imposing the time dependence of $q$ or $p$.

If we wish to have a theory where the conjugate momentum is not the same as the dynamical momentum, then we need to \textit{deconjugate} the momentum $p$ from its dynamically associated $q$~.  As demonstrated above, this requires twice as many variables as those of the standard theory.  For example, in the KvN theory, the need to separate the two types of momenta arises from the desire to formulate classical mechanics in the language of operators.  There, the natural dynamical bracket is the commutator, but maintaining the classical nature of the theory requires observables to commute.  Here we will see that this problem can be formulated in more general terms.

The canonical Hamiltonian theory was defined in Sec.~\ref{sec:summary-hamiltonian}.  Let us now define a deconjugation of that theory.  We shall denote the canonical conjugate of $q$ as $\tilde{p}$ and the canonical conjugate of $p$ as $\tilde{q}$~.  To avoid confusion, we will use the ``tilde-bracket'' $\lb\cdot\,,\cdot\rb_{\!\sim}$ for the new deconjugated relations and keep the original bracket $\lb\cdot\,,\cdot\rb$ for the standard relations.  The deconjugated fundamental relations are
\begin{align}\label{eq:fund-rel-2}
    \lb q,\tilde{p} \rb_{\!\sim} =
    \lb \tilde{q},p \rb_{\!\sim} = 1~, \quad
    \lb q,p \rb_{\!\sim} =
    \lb \tilde{q},\tilde{p} \rb_{\!\sim} = 0~,
\end{align}
and all other brackets vanish.  The number of variables in the theory has doubled.  But this doubling is simply a result of deconjugation or the separation of conjugate ($\tilde{p}$) and dynamical ($p$) momenta, not an addition of truly independent degrees of freedom.

The new fundamental relations~\eqref{eq:fund-rel-2} imply that now $\lb\cdot\,,\tilde{p}\rb_{\!\sim}$ plays the role of $\lb\cdot\,,p\rb$ in the original formulation and similarly for $\tilde{q}$ when acting on $u(q,p)$~:
\begin{align}\label{eq:uqptilde}
    \begin{split}
    \left(\partial_q u\right)_{\!\sim} &= \lb u,\tilde{p} \rb_{\!\sim} = \underline{\lb u,p \rb} = \underline{\partial_q u}~, \\
    \left(\partial_p u\right)_{\!\sim} &= \lb \tilde{q},u \rb_{\!\sim} = \underline{\lb q,u \rb} = \underline{\partial_p u}~.
    \end{split}
\end{align}
We use a tilde subscript to indicate that a quantity is calculated in the tilde-theory.  The fact that $q$ and $p$ commute in the tilde-formulation implies some loss of structure.  We can assume that the tilde-variables ($\tilde{q}$,$\tilde{p}$) encode the missing noncommutative details.  If this is true, then those extra variables might not be unphysical as is usually assumed, though the original theory cannot be uniquely recovered from its deconjugated theory.

The original Hamiltonian $H(q,p)$ cannot give rise to the equations of motion, because its tilde-bracket with $q$ or $p$ vanishes.  We must construct a new tilde-Hamiltonian $\tilde{H}(q,\tilde{p},\tilde{q},p)$ to provide such equations.  If we choose the tilde-Hamiltonian such that $(\dot{q})_{\!\sim}$ is a function of $(q,\tilde{p},\tilde{q},p)$, then it would not be clear which variable plays the role of the dynamical momentum of $q$~.  Thus, the tilde-variables must disappear from the equation of motion for $q$, and similarly for $p$~.  Then the tilde-Hamiltonian must be linear in $\tilde{q}$ and $\tilde{p}$~.  Further, our goal is to merely deconjugate an existing canonical theory, not to introduce another canonical one with twice the degrees of freedom.  Then we ask that the tilde equations of motion for $q$ and $p$ be identical to their original counterparts, save for the commutativity:
\begin{align}\label{eq:qpdot-tilde}
    \begin{split}
    (\dot{q})_{\!\sim} &= \lb q, \tilde{H} \rb_{\!\sim} = \underline{\lb q,H \rb} = \underline{\partial_p H} = \underline{\dot{q}}~, \\
    (\dot{p})_{\!\sim} &= \lb p, \tilde{H} \rb_{\!\sim} = \underline{\lb p,H \rb} = - \underline{\partial_q H} = \underline{\dot{p}}~.
    \end{split}
\end{align}
For these equations to hold, we write the tilde-Hamiltonian as
\begin{align}\label{eq:t-hamiltonian}
    \tilde{H} = \tilde{p} \underline{\partial_p H} + \underline{\partial_q H} \tilde{q} + \underline{\alpha_H(q,p)}~,
\end{align}
where $\alpha_H(q,p)$ is arbitrary.  The placement of $\tilde{q}$ and $\tilde{p}$ in this equation is not unique.  For example, the first term could have been $\underline{\partial_p H} \tilde{p}$.  The choice of ordering and of $\alpha_H$, however, do not affect the equation of motion for any variable $u(q,p)$ as shown below, but will affect those of tilde-variables.

The original Hamiltonian can, in general, be any function of $q$ and $p$.  Therefore, Eqn.~\eqref{eq:t-hamiltonian} defining the tilde-Hamiltonian can be generalized to obtain the ``tilde-image'' of a general original variable $u(q,p)$
\begin{align}\label{eq:t-image}
    \tilde{u}(q,\tilde{p},\tilde{q},p) = \tilde{p} \underline{\partial_p u} + \underline{\partial_q u} \tilde{q} + \underline{\alpha_u(q,p)}~.
\end{align}
The tilde-image is defined up to an arbitrary function $\alpha_u(q,p)$ with the condition $\alpha_q=\alpha_p=0$~.  Taking the tilde-bracket of an original variable $u(q,p)$ with the tilde-image of another variable $\tilde{v}(q,\tilde{p},\tilde{q},p)$, and using~\eqref{eq:uqptilde}, we get
\begin{align}\label{eq:P-like}
    \lb u , \tilde{v} \rb_{\!\sim} 
    = \lb \tilde{u} , v \rb_{\!\sim}
    = \underline{\partial_q u \partial_p v} - \underline{\partial_q v \partial_p u}~.
\end{align}
The tilde-bracket of $u(q,p)$ and $\tilde{v}(q,\tilde{p},\tilde{q},p)$ is just the Poisson bracket of $u(q,p)$ and $v(q,p)$ with commuting $q$ and $p$~.  Equations~\eqref{eq:t-image} and~\eqref{eq:P-like} define the correspondence between the original canonical theory and its deconjugation.

The Poisson bracket appears, and tilde-variables are absent from it.  This is by design.  We chose the tilde-image $\tilde{u}$ of a general variable $u$ in the original theory to be linear in $\tilde{q}$ and $\tilde{p}$ precisely for that to happen.  Creating tilde-images of the canonical theory is, of course, not the only way to have functions of $\tilde{q}$ and $\tilde{p}$~, but we will not be concerned with those.

While the equations of motion for the original variables are independent of $\tilde{q}$ and $\tilde{p}$, the reverse is not generally true.  For example, the simple Hamiltonian
\begin{align}
    H = \frac{p^2}{2m} + V(q)
\end{align}
gives rise to the tilde-Hamiltonian
\begin{align}\label{eq:simple-t-ham}
    \tilde{H} = \frac{\tilde{p}p}{m} + V'(q) \tilde{q} + \underline{\alpha_H(q,p)}~,
\end{align}
and the equations of motion are
\begin{align}
    \dot{q} &= p/m~,\quad
    \dot{p} = -V'(q)~,\\
    \dot{\tilde{q}} &= \tilde{p}/m + \underline{\partial_p \alpha_H}~,\quad
    \dot{\tilde{p}} = -V''(q) \tilde{q} - \underline{\partial_q \alpha_H}~.
\end{align}
This one-sided relationship between the original and the tilde-variables is interesting.  One (visible) set of variables acts on, but is not acted-upon by, the other (hidden) set.  As we will see in Sec.~\ref{sec:no-part-deconj}, this will not remain the case when mixing canonical (quantum) and deconjugated (classical) systems.  Further, since we are restricted to using tilde-Hamiltonians that are linear in $\tilde{q}$ and $\tilde{p}$, we don't have any direct interaction between tilde-variables.

For completeness, we can find the tilde-Lagrangian given by a Legendre transform of the tilde-Hamiltonian
\begin{align}\label{eq:t-lagrnagian}
    \tilde{L} = \tilde{p}\dot{q} + p\dot{\tilde{q}} - \tilde{H}~.
\end{align}
For the simple tilde-Hamiltonian~\eqref{eq:simple-t-ham}, we get
\begin{align}
    \tilde{L} = m\dot{q}\dot{\tilde{q}} - V'(q)\tilde{q} - \alpha_H(q,p)~.
\end{align}

Finally, a note on conserved quantities after deconjugation.  From the equations of motion for $q$ and $p$~\eqref{eq:qpdot-tilde}, we see immediately that if $p$ (or $q$) is conserved in the original canonical theory, then it is also conserved (under $\tilde{H}$) in the deconjugated one.  Further, it is easy to show that $\tilde{p}$, too, will be conserved for a certain choice of $\alpha_H$ in $\tilde{H}$~.  To generalize the relationship between conserved quantities in the canonical and the deconjugated theories, we make use of McCoy's formula~\cite{McCoy1929}
\begin{align}\label{eq:McCoy}
    \lb u,v \rb =  \sum_{k=1}^\infty \frac{(-i\hbar)^{k-1}}{k!} \left( \partial_q^k u \partial_p^k v - \partial_q^k v \partial_p^k u \right)~.
\end{align}
If $\lb u,H \rb = 0$, then underlining (setting $qp=pq$) the terms in the expansion, we get
\begin{align}\label{eq:h-expansion}
    \sum_{k=1}^\infty \frac{(-i\hbar)^{k-1}}{k!} \left( \underline{\partial_q^k u \partial_p^k H } - \underline{\partial_q^k H \partial_p^k u} \right) = 0.
\end{align}
If $u$ and $H$ do not contain $\hbar$ explicitly, the underlined terms do not contain $\hbar$'s at all since $q$ and $p$ are set to commute.  Then each term in the expansion~\eqref{eq:h-expansion} must vanish individually.  Noticing that the first term is nothing but the tilde-bracket of $u$ and $\tilde{H}$, we get the result
\begin{align}
    \lb u,H \rb = 0 \implies \lb u,\tilde{H} \rb_{\!\sim} = 0~.
\end{align}
The reverse is generally not true
\begin{align}
    \lb u,\tilde{H} \rb_{\!\sim} = 0 \centernot\implies \lb u,H \rb = 0~.
\end{align}
Thus, conservation in the canonical theory implies conservation in its deconjugation, but not vice versa\footnote{In fact, any generator $G$ of transformation in the canonical theory will have its role played by its tilde-image $\tilde{G}$ in the deconjugated theory.  This emphasises that tilde-images are not truly ``new'' or separate variables.}.  This is expected if the canonical theory is quantum and the deconjugated one represents classical mechanics.

The general arguments of this section are valid for systems of any number of degrees of freedom as long as \textit{all} degrees of freedom are deconjugated.  In Sec.~\ref{sec:no-part-deconj} we shall see that deconjugating only a subset of the degrees of freedom is, in general, problematic.

\section{KvN is deconjugated QM}
\label{sec:kvn}

The Koopman-von Neumann operator formulation of classical mechanics~\cite{koopman1931,bondar2012,wilczek2015} uses the same tilde-commutation relations~\eqref{eq:fund-rel-2}~.  Indeed, the generator of time evolution~\eqref{eq:t-hamiltonian} obtained here appears in the KvN literature.  The deconjugation view presented here allows us to see KvN not only as a formulation of classical mechanics, but as one extracted directly from quantum mechanics.  Looking at McCoy's formula~\eqref{eq:McCoy}, we see that the tilde-bracket of any variable $u(q,p)$ and the tilde-image of any other variable $v(q,p)$ is the same as the underlined $\hbar^0$ term in the expansion for the quantum bracket $\lb u,v \rb$~.  In other words, the KvN bracket can be obtained by setting the commutator of $q$ and $p$ to zero (underlining) then taking $\hbar\to0$.  This is reminiscent of phase-space quantum mechanics viewed as a deformation of classical mechanics.

Let us examine this further.  The KvN formulation describes classical mechanics in terms of operators, often called the ``language of quantum mechanics''.  It is sometimes compared to the phase-space formulation of quantum mechanics which uses the ``language of classical mechanics'', functions on phase space.  This analogy, however, might be superficial.  While phase-space QM takes simple CM and deforms it until it becomes quantum, an operator description of a commutative theory (classical mechanics) requires additional complications: the inclusion of tilde-variables.

Deconjugation (KvN) finds classical equations of motion and uncertainty relations out of quantum ones, but at the cost of introducing new quantum equations and uncertainties:
\begin{align}\label{eq:t-uncertainty}
    \begin{split}
    \Delta q \Delta \tilde{p} &\geq \hbar / 2~,\\
    \Delta \tilde{q} \Delta p &\geq \hbar / 2~.
    \end{split}
\end{align}
A rich quantum structure still exists.  The tilde-variables are not easily dismissed as unphysical, they generate transformations and can be conserved quantities as mentioned in the previous section.  In the next section, we show that the tilde-variables do not remain ``hidden'' when a KvN system interacts with a quantum one.

\section{Partial deconjugation and Q-C systems}
\label{sec:no-part-deconj}

We have seen that the process of deconjugation alters the canonical relations yet reproduces the original equations of motion modulo the ordering of $q$ and $p$.  This is true when the process is applied to \textit{all} degrees of freedom.  However, the equations of motion will be altered when deconjugating only a subset of the degrees of freedom.  As we discuss in this section, this might affect the use of the KvN theory in the study of quantum-classical interactions.  For a treatment of quantum-classical systems using KvN theory, see~\cite{bondar2019} and references therein.

Consider a system of two degrees of freedom $q_1$ and $q_2$ with canonical momenta $p_1$ and $p_2$.  The equations of motion in the standard formulation with a Hamiltonian $H(q_i,p_i)$ are
\begin{align}
    \dot{q}_i = \lb q_i,H \rb~,\quad \dot{p}_i = \lb p_i,H \rb~,
\end{align}
where $i=1,2$~.  Now suppose that we choose to deconjugate only $q_1$ and $p_1$, while keeping the standard canonical relation between $q_2$ and $p_2$ intact.  The new canonical relations are
\begin{align}\label{eq:part-fund-rel}
\begin{split}
    \lb q_1,\tilde{p}_1 \rb_{\!\sim} = \lb \tilde{q}_1,p_1 \rb_{\!\sim} = \lb q_2,p_2 \rb_{\!\sim} &= 1~,\\
    \lb q_1,p_1 \rb_{\!\sim} &= 0~.
\end{split}
\end{align}
All other brackets vanish.  Following the procedure in Sec.~\ref{sec:deconj}, we set the tilde-Hamiltonian to 
\begin{align}
    \tilde{H} = \tilde{p}_1 \underline{\partial_{p_1}H} + \underline{\partial_{q_1}H} \tilde{q}_1 + \underline{H} + \underline{\alpha_H(q_1,p_1)}~.
\end{align}
Here the underlines mark expressions in which $q_1$ and $p_1$ are set to commute.  This tilde-Hamiltonian gives the correct equations of motion for $q_1$ and $p_1$:
\begin{align}
    (\dot{q}_1)_{\!\sim} = \underline{\partial_{p_1}H}~, \quad (\dot{p}_1)_{\!\sim} = - \underline{\partial_{q_1}H}~.
\end{align}

However, for $q_2$ and $p_2$ (the still-conjugated part of the system) we get
\begin{align}
    \begin{split}
        (\dot{q}_2)_{\!\sim} &= \lb q_2,\tilde{H} \rb_{\!\sim} \\
        &= \lb q_2,\underline{H} \rb + \tilde{p}_1 \lb q_2, \underline{\partial_{p_1}H} \rb + \lb q_2, \underline{\partial_{q_1}H} \rb \, \tilde{q}_1~,\\
    \end{split}
\end{align}
and
\begin{align}
    \begin{split}
        (\dot{p}_2)_{\!\sim} &= \lb p_2,\tilde{H} \rb_{\!\sim} \\
        &= \lb p_2,\underline{H} \rb + \tilde{p}_1 \lb p_2, \underline{\partial_{p_1}H} \rb + \lb p_2, \underline{\partial_{q_1}H} \rb \, \tilde{q}_1~.
    \end{split}
\end{align}
We get extra terms, proportional to $\tilde{q}_1$ and $\tilde{p}_1$, in the equations of motion for $q_2$ and $p_2$.  Preserving the original equations of motion for the deconjugated part of the system spoils those of the non-deconjugated part.

Before discussing the implications of this, we can first investigate the possibility of restoring the correct equations of motion dynamically; that is, through the equations of motion for $\tilde{q}_1$ and $\tilde{p}_1$.  Consider the simple Hamiltonian
\begin{align}
	H = \frac{p_1^2}{2m_1} + \frac{p_2^2}{2m_2} + V(q_1,q_2)~.
\end{align}
The tilde-Hamiltonian then is
\begin{align}
	\tilde{H} = \frac{\tilde{p}_1p_1}{m_1} + \frac{p_2^2}{2m_2} + \partial_{q_1} V + V + \alpha_H(q_1,p_1)~.
\end{align}
The equations of motion for $q_2$ and $p_2$ become
\begin{align}
	(\dot{q}_2)_{\!\sim} = \frac{p_2}{m_2}~, \quad (\dot{p}_2)_{\!\sim} = - \partial_{q_2} V - \partial_{q_2} \partial_{q_1} V \, \tilde{q}_1~.
\end{align}
We see that if the ``unphysical'' $\tilde{q}_1$ is dynamically zero for all times, the original equations of motion for $p_2$ are  recovered.  For the simple Hamiltonian above, this is possible since the equation of motion for $\tilde{q}_1$ combined with that of $\tilde{p}_1$ gives
\begin{align}
	\ddot{\tilde{q}}_1 = \frac{1}{m_1} \left( -\partial_{q_1}^2 V \, \tilde{q}_1 + \dot{p}_1 - \partial_{q_1} \alpha_H \right) + \frac{d}{dt} \partial_{p_1} \alpha_H~.
\end{align}
We can set the arbitrary function to be $\alpha_H=-p_1^2/2m_1$, which gives
\begin{align}
	\ddot{\tilde{q}}_1 = -\frac{1}{m_1} \partial_{q_2}^2 V \, \tilde{q}_1~.
\end{align}
Now, if we assert that $\tilde{q}_1$ is an unphysical variable, then we can choose its initial conditions so that $\tilde{q}_1(t)=0$.  This choice restores the original equation of motion for $p_2$.

Treating $\tilde{q}_1$ and $\tilde{p}_1$ this way may be problematic, however.  As mentioned before, even though they are not the physical position and momentum, they are still the generators of translation in the momentum and configuration spaces,  respectively.  Further, should we think of those variables as unphysical if the choice of their initial conditions affects the physical degrees of freedom?  Even if we choose to ignore these conceptual problems, the trick used above is not possible for more general Hamiltonians where the potential is a function $V(q_1,p_1,q_2,p_2)$ of both positions and momenta.

It is possible that the role of the auxiliary variables in a mixed system becomes obscured in the Schrödinger picture, but analyzing the problem in the Heisenberg picture makes it clear.  The issue needs to be confronted.

The auxiliary variables surface when KvN and quantum systems are mixed.  But this happens more generally, whenever deconjugated and canonical systems are combined.  For example, part of a classical system can be deconjugated such that we have the fundamental relations~\eqref{eq:part-fund-rel},  with quantum brackets replaced by Poisson brackets.  Partial deconjugation makes auxiliary variables appear in the physical equations of motion even for a classical-classical system.

The difficulties discussed in this section should be considered when studying quantum-classical dynamics using the KvN formulation.  The classical part of the system is a deconjugated part while the quantum remains canonical.  Even though the ``original'' equations of motion for the system are unknown, we see from this section that the KvN path to quantum-classical equations of motion might not be straightforward.

\section*{Conclusion}

The no-go theorems for quantum-classical mixing  (see~\cite{gil2017} and references therein) reveal a clash between the canonical structures of classical and quantum mechanics.  Although this incompatibility can be circumvented for certain classes of systems~\cite{AminQCDB2021,AminIllustration2021}, it is recognized that the mixing is not generally  possible.

We do not consider here the option of mixing canonical quantum mechanics with a theory that is not canonical. Selecting the kind of non-canonical system is  difficult, as is the  crucial task of showing it has a strong relation to classical mechanics. 

The impediment to canonical hybrid quantum-classical systems is the failure of the quantum-classical bracket to satisfy the Leibniz and Jacobi properties necessary for consistent canonical dynamics.  At first glance, the Koopman-von Neumann approach  seems to bypass the inconsistency issue of the dynamical bracket. It uses the commutator, which automatically satisfies the required conditions.

However, we have shown here that the KvN theory is  \textit{deconjugated} canonical quantum theory.  Classical equations of motion do appear, but at the cost of introducing auxiliary variables. KvN theory is more than just an operator formulation of classical mechanics. It is not physically clear that the auxiliary variables can be ignored.  

Even if it may be  possible to argue away the auxiliary variables as unphysical for a single system, they are indispensable in a hybrid quantum-KvN system (see also   \cite{wilczek2015}).  

We see that the mixing of classical and quantum systems remains somewhat problematic even in an operator formulation of both systems. Using the operator commutator for both quantum and classical systems raises new issues.

\section*{Acknowledgements}
This research was supported in part by a Discovery Grant (M.A.W.) from the Natural Sciences and Engineering Research Council of Canada (funding Reference No. RGPIN-2022-04225). 

\footnotesize \bibliography{references}

\end{document}